\documentclass[final,3p,times,twocolumn]{elsarticle}

\usepackage{amssymb,amsmath}

\journal{Nuclear Instruments and Methods B}

\begin{document}

\begin{frontmatter}



\title{Matching Optimization for TRIUMF's Rare Isotope Linac}
\author[inst1,inst2]{O. Shelbaya}

\affiliation[inst1]{organization={TRIUMF},
            addressline={4004 Wesbrook Mall}, 
            city={Vancouver},
            postcode={V6T 2A3}, 
            state={BC},
            country={Canada}}
\author[inst1,inst2]{O.~Hassan}
\author[inst1,inst2]{R.~Baartman}%
\author[inst1,inst2]{O.~Kester}
\author[inst1]{C.~Pearce}
\author[inst1,inst2]{T.~Planche}
\author[inst1,inst2]{L. Zhang}

\affiliation[inst2]{organization={Department of Physics and Astronomy, University of Victoria},
            addressline={PO Box 1700 STN CSC}, 
            city={Victoria},
            postcode={V8W 2Y2}, 
            state={BC},
            country={Canada}}

\begin{abstract}
Fringe field effects are computed from simulated and measured quadrupole magnetic field gradients, enabling a more realistic modelling of the medium energy section optics in TRIUMF's RIB postaccelerator. Quantifying the fringe field optics using the method developed by Matsuda and Wollnik allows for the use of parallel modelling tune optimizations for machine operation, providing operators with an efficient means to re-tune the variable energy drift tube linac for RIB delivery. 
\end{abstract}




\end{frontmatter}



\section{\label{sec:introduction}Introduction}

\begin{figure}[!b]
    \centering
    \includegraphics[width=0.35\textwidth]{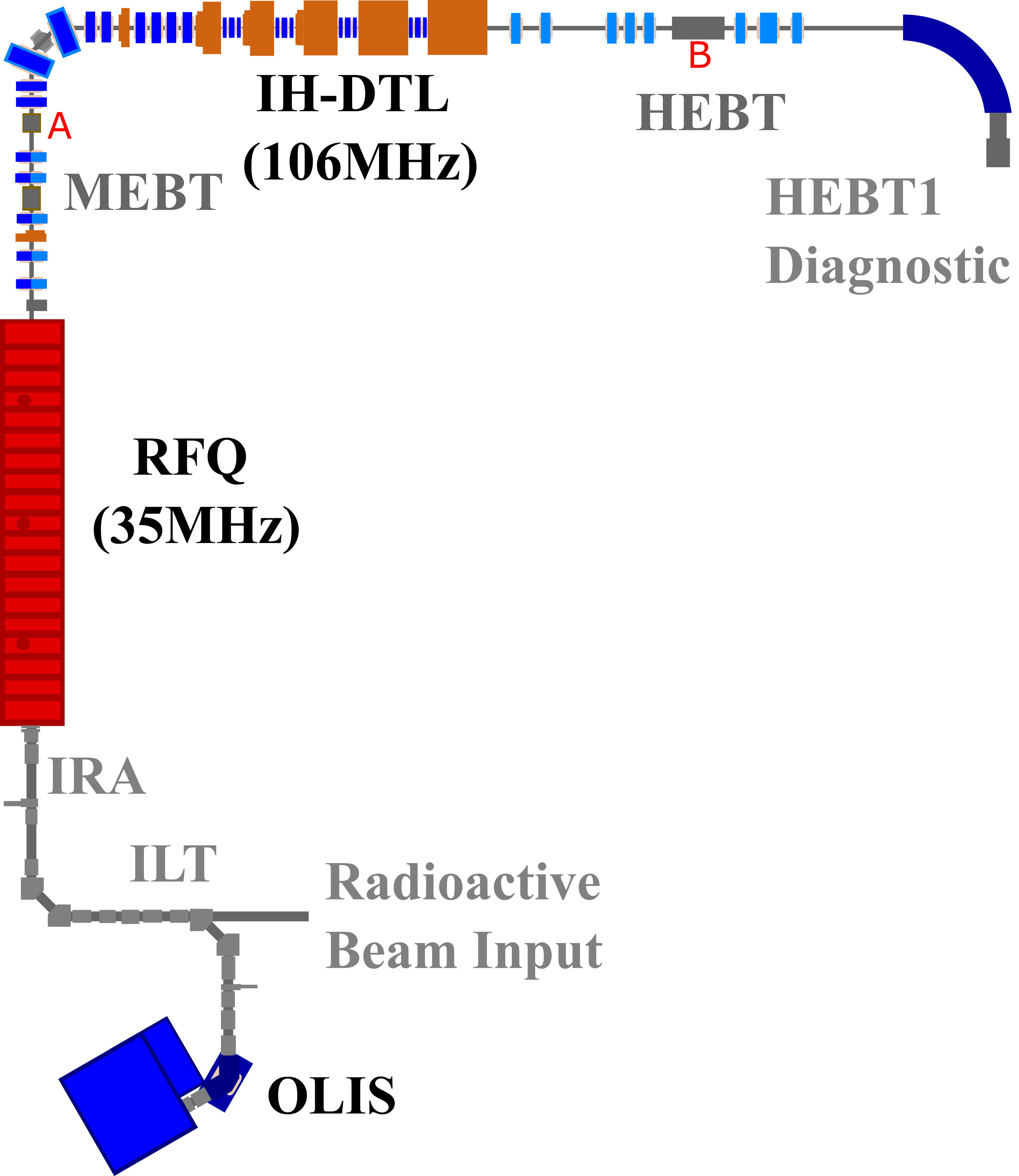}
    \caption{Overview of the ISAC linac, including RFQ and DTL and the MEBT section. A 90$^\circ$ bend allows for charge state selection, if a retractable stripping foil is inserted at location A. Another post-DTL foil can be used at location B.}
    \label{fig:MEBTOverview}
\end{figure}

At TRIUMF, the Isotope Separator and Accelerator (ISAC) facility uses proton beams, driven by the 18\,m diameter 520\,MeV multi-user cyclotron to produce rare isotopes spanning the nuclear chart\cite{kunz2014nuclear}. Postacceleration is achieved using a paired rf quadrupole (RFQ) and drift tube linac (DTL) (Figure \ref{fig:MEBTOverview}) to deliver beams to experimental end-users. By design, the RFQ accepts low energy beams up to $A/q$\,=\,30, while DTL operates in the 2\,$\leq$\,$A/q$\,$\leq$\,6 region. The DTL is a separated function\cite{Laxdal:1997ge} machine, itself producing fully variable output energy beams, meaning the accelerator must also constantly be tuned for different requested beam compositions and energies.

\begin{figure*}[!t]
    \centering    \includegraphics[width=\textwidth]{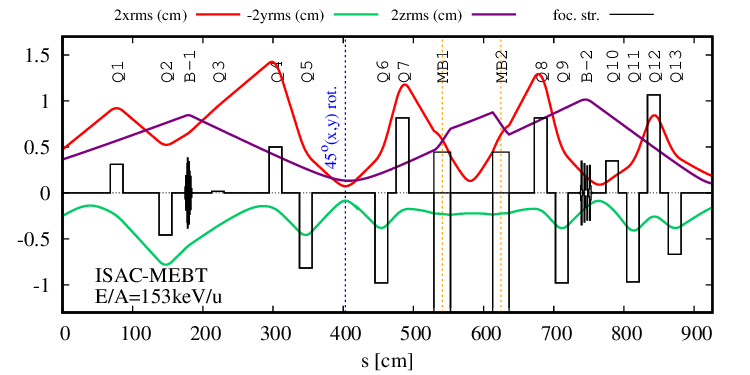}
    \caption{\label{fig:mebt_design_tune}Design MEBT tune showcasing 2rms envelopes for ($x,y,z$) computed in the envelope code {\tt TRANSOPTR}. The 45$^\circ$ dipoles are marked MB1 and MB2. In this configuration, a longitudinal ($z$) focus is established at the stripping foil using the first rf buncher, with the second buncher focusing into the first DTL tank. Normalized transverse focal strength for quadrupoles, dipoles and RF cavities shown in black.}
\end{figure*}
We consider the ISAC-MEBT optics~\cite{TRI-BN-19-02, laxdal2002beam}, whose design tune is shown in Figure~\ref{fig:mebt_design_tune}. The MEBT section accepts accelerated RFQ beam and shapes it for injection into DTL\cite{TRI-BN-22-29}. Operators have had to manually tune the MEBT quadrupoles to match beam for acceleration. This process is both time consuming and difficult to train for, imposing additional complexity and overhead on machine tuning. Specifically, the tuning approach is being transitioned from a subjective, manual process to a more objective methodology guided by concurrent beamline modelling.

This work was motivated by the desire to use previously reported\cite{shelbaya2021autofocus} parallel modelling and tune optimization software, enabling operators to efficiently control the optics, obtaining high transmission without a necessity to manually tune quadrupoles for transmission. As TRIUMF transitions into the ARIEL era\cite{merminga2011ariel,Bagger:IPAC2018-MOXGB2}, with an anticipated increase of delivered radioactive beam-hours\cite{dilling2013ariel}, reduction of overhead tuning times is critical if increased accelerated beam production is to be successfully achieved.

As presented in this work, the omission of fringe field effects in beam optic simulations and tune computations of MEBT results in a transverse mismatch at DTL injection. By incorporating fringe field lensing effects to our envelope model, we are able to compute tunes which achieve injection into DTL and high-transmission, without the need for quadrupole gradient fine-tuning. This improves tuning efficiency by reducing the linac optics' configuration space and paves the way to autonomous tuning algorithms, presently under development at TRIUMF\cite{wang2021accelerator,ghelfi2025bayesian,shelbaya2024tuning} to support operators during accelerated RIB delivery.

\subsection{\label{sec:MEBTSECTION}The ISAC-MEBT Tune}

The MEBT section comprises thirteen magnetic quadrupoles, two 45$^\circ$ bending dipoles, and two RF bunching cavities~\cite{marchetto2014high}. It captures the divergent output of the RFQ at an $E/A = 0.153$\,MeV/u and prepares the beam for injection into the downstream DTL. The corresponding design optics are illustrated in Figure~\ref{fig:mebt_design_tune}. An RF chopper is installed between quadrupoles Q3 and Q4 to remove two of the three RF buckets exiting the RFQ, ensuring appropriate bunch selection for acceleration~\cite{shelbaya2019fast}.

The transverse optics are configured to produce a round beam spot at the location of the movable stripping foil (Fig.\,\ref{fig:mebt_design_tune}, $s = 400$\,cm), with a 1:1 aspect ratio in both the ($x,x'$) and ($y,y'$) phase spaces. This is achieved using the first five quadrupoles of the MEBT section~\cite{laxdal2014isac}, which are rotated by 45$^\circ$ to match the orientation of the RFQ vanes. At the foil, the beamline reference frame is rotated back by 45$^\circ$, allowing all downstream quadrupoles and steering elements to be mounted in the standard laboratory ($x,y$) frame and orientation. A 2\,mm slit located near the foil imposes a constraint on the horizontal beam size~\cite{marchetto2010isac}. In addition, the first RF bunching cavity (Fig.\,\ref{fig:mebt_design_tune}, labeled B-1) is employed to minimize the temporal spread of the beam at the foil location, to minimize emittance growth from foil scattering when in use\cite{laxdal1997separated}.

\section{B-I Calibration\label{sec:BICalib}}

\begin{figure}[!t]
\centering
\includegraphics[width=\linewidth]{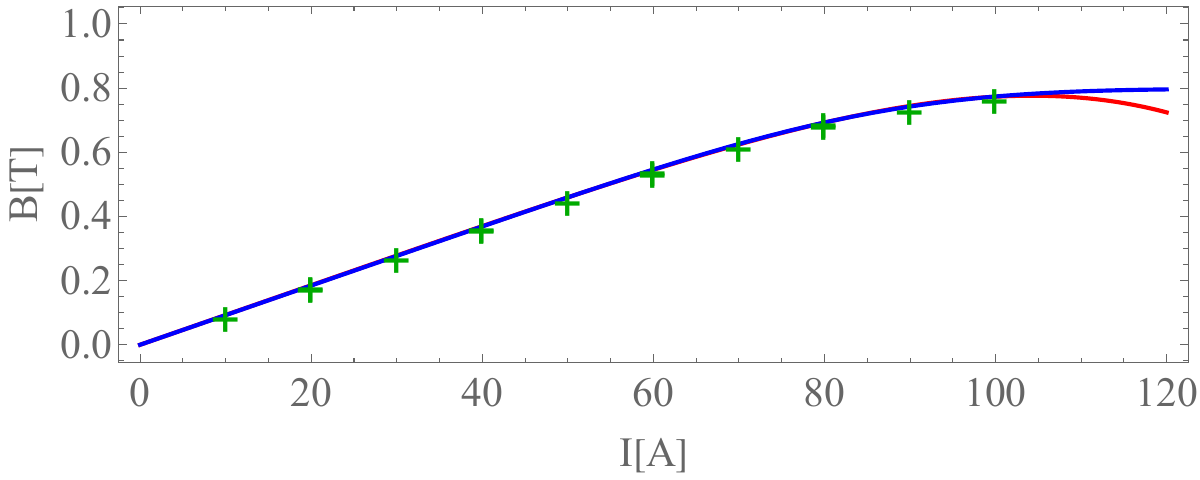}
\caption{\label{fig:BI_Danfysik_L1_Evans}Measured $B$-$I$ calibration curve with pseudo-Langevin fit (blue) for the {\it Danfysik L1} quadrupole (data in green). The polynomial 3-parameter fit is shown (red) for comparison.} 
\end{figure}

Magnetic survey measurements were performed on the MEBT {\it Danfysik L1} (1987) type quadrupoles, and the resulting $B$-$I$ relationship was fitted using a modified pseudo-Langevin function, consistent with the methodology employed in~\cite{TRI-BN-19-18}, which has only two fit parameters:
\begin{equation}
B(I) = \frac{a_1}{a_3}\tanh\left[(a_3I) + \frac{1}{3}(a_3I)^3 + \frac{1}{5}(a_3I)^5 \right].
\label{eq:langevin}
\end{equation}
The fitting parameters, summarized in Table~\ref{tab:BIMEBTQuad}, provide an accurate empirical model for use in beam optics simulations. The fitted curve and corresponding residuals are shown in Figure~\ref{fig:BI_Danfysik_L1_Evans}, demonstrating sub-percent agreement across the relevant current range.
\begin{table}[!t]
\centering
\caption{\label{tab:BIMEBTQuad} Fit parameters for the pseudo-Langevin $B$--$I$ calibration curve of the {\it Danfysik L1} quadrupole.} 
\vspace{0.5em}
\begin{tabular}{l c}
\hline
Parameter & Value \\
\hline
$a_1$ [T/A]         & $(9.23 \pm 0.03) \times 10^{-3}$ \\
$a_3$ [1/A]         & $(1.156 \pm 0.007) \times 10^{-2}$ \\
\hline
\end{tabular}
\end{table}

The pseudo-Langevin form captures the expected magnetic saturation using only two parameters. This provides a compact and robust model that better reflects the magnet's physical characteristics. As current increases toward saturation, a polynomial fit requires progressively more parameters to accurately describe the behaviour, and still lacks the physically correct asymptotic limit. In contrast, the pseudo-Langevin function provides reliable extrapolation and remains consistent with the magnet's nonlinear response over the full current range and beyond.

\section{Opera3D Simulations}

\begin{figure}[!htpb]
  \centering
  \includegraphics[width=\linewidth, trim={13cm 2.5cm 13cm 3cm}, clip]{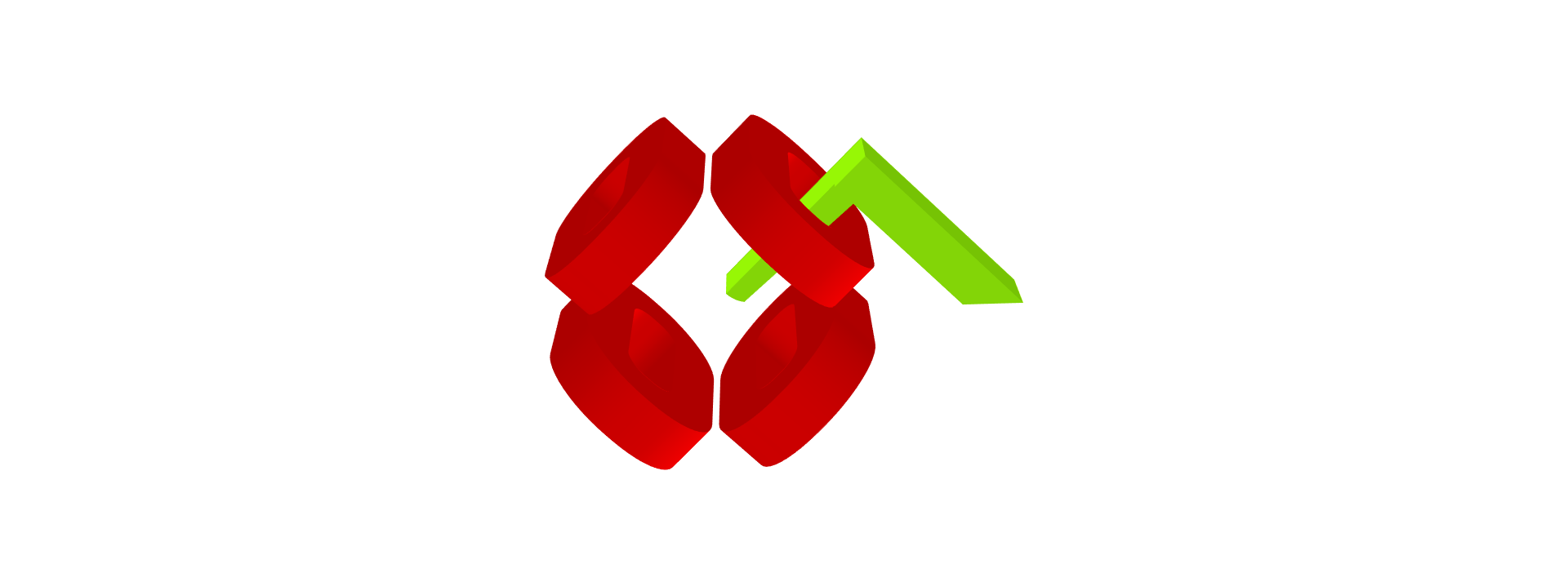}
  \caption{{\tt Opera-3D} 1/16 model of Danfysik L1 magnet, which preserves all symmetries of the quadrupole. The magnet yoke's physical aperture radius is 2.6\,cm.}
  \label{fig:operaQuad}
\end{figure}

A detailed electromagnetic model of the Danfysik L1-type quadrupoles was developed in Opera-3D (Fig.~\ref{fig:operaQuad}), based on the corresponding mechanical geometry imported from a SolidWorks design. The magnetic field gradient of the resulting model was analyzed to determine both the effective magnetic length, $L_{\rm eff}$, and to characterize the fringe field region, with particular attention to its optical properties.

To evaluate the focusing properties of the quadrupole, {\tt TRANSOPTR}\cite{heighway1981transoptr,TRI-BN-22-08} was employed. Using its built-in RK integrator with an adaptive step-size, the effect of the quadrupole is computed by integration of the longitudinal gradient profile upon the beam, performed with a user-specified relative error tolerance of 10$^{-4}$. This approach provides a continuous simulation of the quadrupole, enabling direct computation of the transfer matrix elements from the gradient distribution. These elements were then used to define an equivalent hard-edge quadrupole representation, characterized by the effective length $L_{\rm eff}$ and an effective aperture radius $a$, which controls the gradient and therefore strength of the quadrupole.

A comparison of the updated hard-edge parameters with the previously used values is presented in Table~\ref{tab:quadfits}. Although the differences in the hard-edge parameters are modest, they result in non-negligible changes to the overall lattice performance, particularly due to effects discussed in the preceding section.

\begin{table*}[htbp]
    \centering
    \caption{Effective quadrupole parameters and corresponding Wollnik integrals for each model.}
    \label{tab:quadfits}
    \begin{tabular}{lccccc}
        \hline
        \textbf{Case} &
        \boldmath\( L_{\mathrm{eff}} \) \textbf{[cm]} &
        \boldmath\( a_0 \) \textbf{[cm]} &
        \boldmath\( I_1 \) &
        \boldmath\( I_2 \) &
        \boldmath\( I_3 \) \\
        \hline
        Original {\tt Trace-3D} model           & 18.00 & 2.60 & -     & -      & -     \\
        Rectangular survey (r = 0.75")          & 17.53 & 2.53 & 0.484 & -0.297 & 0.190 \\
        Cylindrical survey (r = 0.75")          & 17.89 & 2.59 & 0.427 & -0.239 & 0.151 \\
        {\tt Opera-3D} (r $\approx$ 0)          & 17.69 & 2.56 & 0.368 & -0.222 & 0.116 \\
        \hline
    \end{tabular}
\end{table*}

\subsection{Hard-edge Model}

A two-step process is followed. First, the gradient distribution is supplied as an input file to {\tt TRANSOPTR}, and the integrated transfer matrix is computed through the field and for a defined drift length downstream of the quadrupole.

In the second step, a simulation is performed using a hard-edge quadrupole model combined with entrance and exit fringe fields whose optics are quantified with the Wollnik integrals. The hard-edge quadrupole and the drift are arranged such that their total physical length matches that of the gradient distribution plus the downstream drift used in the first simulation.

A fit routine in {\tt TRANSOPTR} is then used to determine the effective hard-edge length and aperture radius that reproduce the same transfer matrix elements. The fitting uses the known $B$--$I$ calibration discussed in Section~\ref{sec:BICalib}.

\subsection{Fringe Field Effects}

Matsuda and Wollnik characterize the quadrupole fringe region interaction using four integrals\cite{matsuda1972third,baartman2007short}:
\begin{align}
    I_1 &= \int_{-\infty}^{s_{\infty}}\int_{-\infty}^{s}k(s'){\rm d}s'{\rm d}s - \frac{s^2}{2}
    \label{eq:wollnik1}\\
    I_2 &= \int_{-\infty}^{s_{\infty}}\int_{-\infty}^{s}k(s'){\rm d}s'{\rm d}s - \frac{s^3}{3}
    \label{eq:wollnik2}\\
    I_3 &= \int_{-\infty}^{s_{\infty}}\Bigg(\int_{-\infty}^{s}k(s'){\rm d}s'\Bigg)^2{\rm d}s\ - \frac{s^3}{3}
    \label{eq:wollnik3}\\
    I_4 &= \int_{-\infty}^{s_{\infty}}k(s)^2{\rm d}s - s_{\infty}.
    \label{eq:wollnik4}
\end{align}
Here $s$ is the length along the beam's trajectory, $s_{\infty}$ is a position within the quadrupole, where the field strength has reached a plateau and $k(s)$ is the quadrupole strength. These provide a quantitative description\cite{matsuda1972third,baartman2007short} of the fringe optics, where the field transitions from its saturation value within the quadrupole to zero in the far-field region. For first-order optics calculations, as implemented in {\tt TRANSOPTR}, only the first three integrals ($I_1$ to $I_3$) are required; $I_4$ is included for completeness; it is used to correct the second order aberration. 

\begin{figure}[!htpb]
  \centering
  \includegraphics[width=\linewidth]{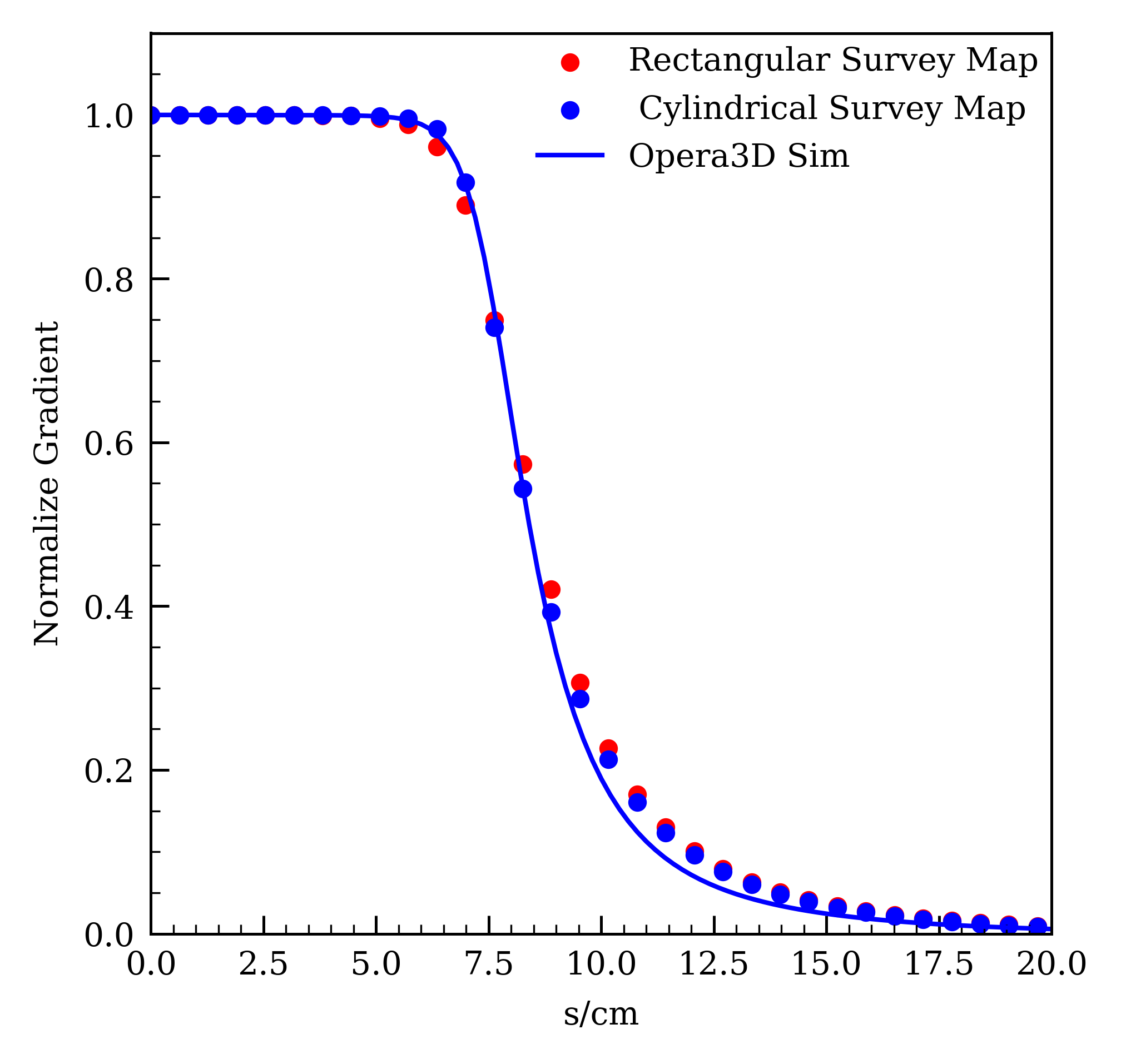}
  \caption{
    Axial quadrupole field gradient profiles for three models,
    illustrating differences in the fringe field regions.
    All curves are plotted on the same axes for direct comparison.
    The central region shows agreement across models, while the gradient
    falloff beyond the effective length varies depending on the measurement
    or simulation method. See Table~\ref{tab:quadfits} for model descriptions.
  }
  \label{fig:quadFringeCompare}
\end{figure}

Three separate datasets are considered: (1) a rectangular and (2) cyclindrical survey performed prior to machine commissioning in the late 90's and (3) a simulated gradient profile in {\tt Opera-3D} at r$\approx$0. The integrals are listed in Table \ref{tab:quadfits} for these cases which are shown in Figure~\ref{fig:quadFringeCompare}. The fringe field integrals are found to differ for near- and off-axis samplings. This radial dependence, is beyond a first order treatment. 

To evaluate whether or not these differences affect the tune to first order, consider the infinitesimal transformation of the beam distribution over a small propagation increment ${\rm d}s$. The transfer matrix $T = I-{\bf F}{\rm d}s$, where $I$ is the identity. The force matrix ${\bf F}$ depends on the second derivatives of the hamiltonian\cite{shelbaya2019fast} and for a single fringe field transit are:
\begin{equation}
    {\bf F_{x}} = 
    \begin{pmatrix}
    1 \mp \xi I_1 & -2\xi aI_2\\
    -\xi^2I_3/a & 1 \pm \xi I_1
    \end{pmatrix}
\label{eq:wollnik_fmatrix_x}
\end{equation}
and
\begin{equation}
    {\bf F_{y}} = 
    \begin{pmatrix}
    1 \pm \xi I_1 & 2\xi aI_2\\
    -\xi^2I_3/a & 1 \mp \xi I_1
    \end{pmatrix},
\label{eq:wollnik_fmatrix_y}
\end{equation}
with the parameter:
\begin{equation}
    \xi = \frac{a(B\rho)}{B_0L_{\rm eff}}
    \label{eq:wollnik_xi}
\end{equation}
where ($B\rho$) is the rigidity and $B_0$ the field at the magnet pole-tip. The parameter $\xi$ quantifies the magnitude of the fringe optics; when it is zero, ${\bf F_{x}}$ and ${\bf F_{y}}$ reduce to the identity matrices and there are no fringe field effects.

\subsection{Quad-Steerer Test}

A beam-based method was used to determine the effective magnetic length of a quadrupole in the MEBT section. A transverse steerer upstream of the quadrupole was used to introduce a known kick, while all other downstream quadrupoles were turned off, leaving only the quadrupole under study powered. The beam was tracked to a downstream profile monitor, where the transverse beam centroids and sizes were recorded. The condition of interest corresponds to the quadrupole setpoint at which the transverse kick to the beam when it is off-centre in the quadrupole exactly cancels the applied steering kick, resulting in zero net displacement of the beam centroid at the monitor. Under this condition, scanning the steerer current produces no observable change in centroid position, indicating that the quadrupole’s effective length matches that required to counteract the kick.

\begin{figure}[h!]
    \centering
    \includegraphics[width=\linewidth]{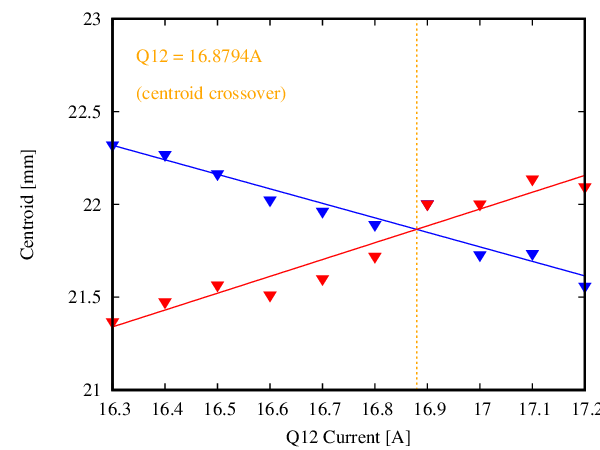}\\
    \includegraphics[width=\linewidth]{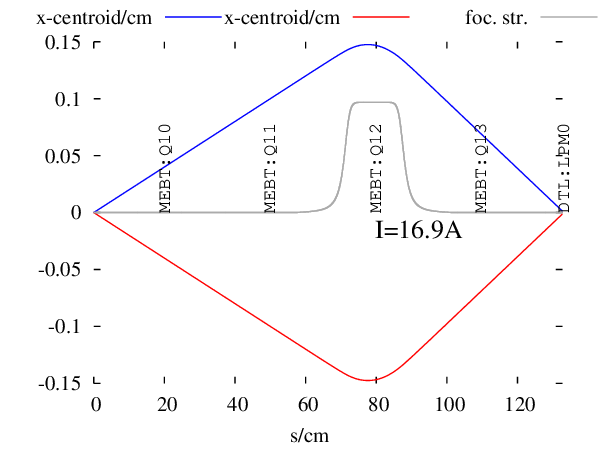}
    \caption{Beam centroids simulated using {\tt TRANSOPTR}, showing the compensation of the steerer kick by the effective dipole fields of the quadrupoles.}
    \label{fig:opera_model}
\end{figure}

The {\tt Opera-3D} field is used with {\tt TRANSOPTR}. The simulated centroids for two steerer settings are shown in Figure~\ref{fig:opera_model}, representing the observed condition on the profile monitor. This agreement between the beam-based measurement and the numerical model validates the simulated gradient distribution and B-I calibration applied to it. Each of the magnetic gradient distributions listed in Table~\ref{tab:quadfits}, including the 18\,cm hard-edge approximation, yields beam centroid predictions consistent with on-line measurements. Variations in effective length across these distributions result in centroid shifts at the downstream profile monitor of less than 0.1\,mm.

\subsection{Quadrupole Interference}

\begin{figure}[!b]
    \centering
    \includegraphics[width=0.85\linewidth]{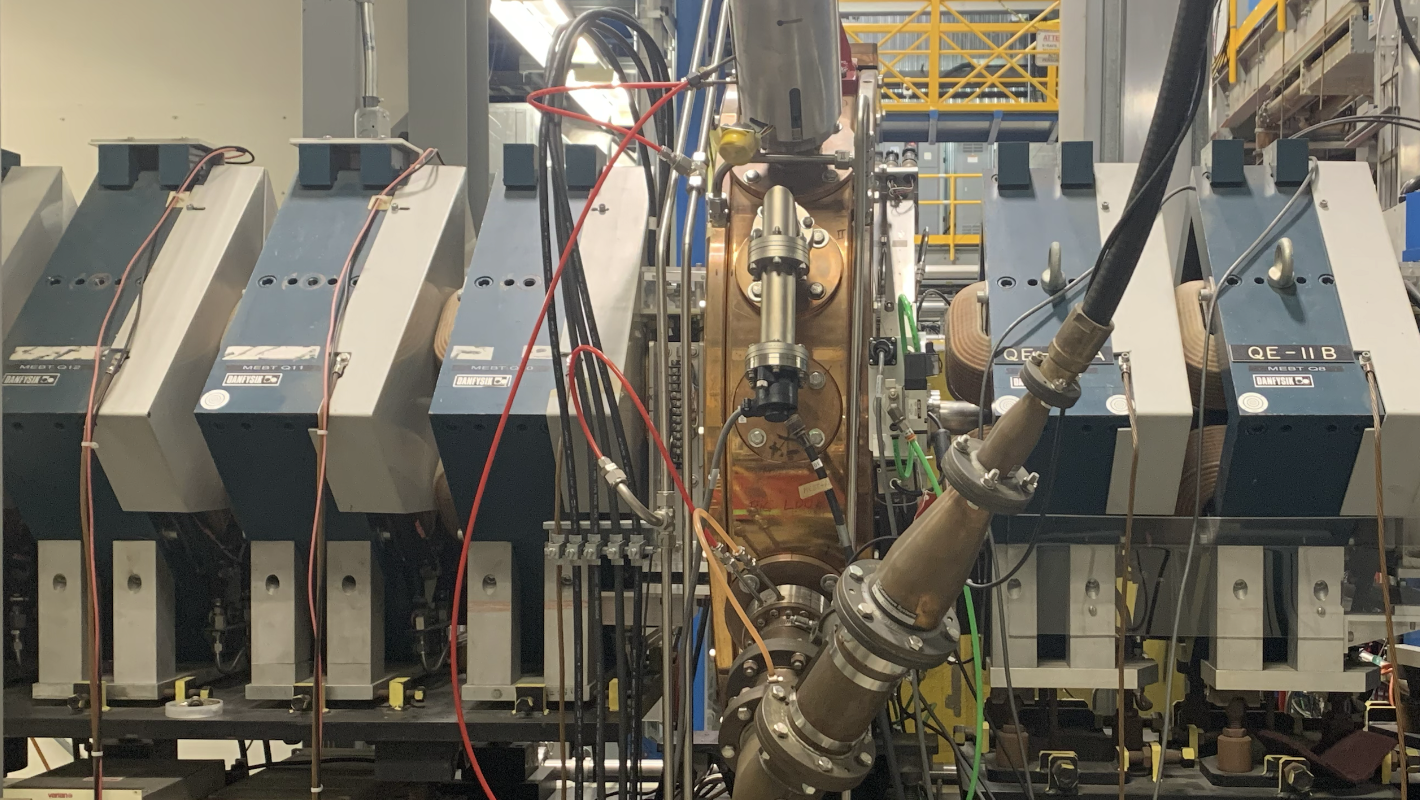}
    \caption{Image of the MEBT section, showing the close spacing of quadrupole magnets. Despite field overlap, no interference effects are observed in simulations using {\tt Opera-3D}.}
    \label{fig:mebt_lattice_spacing}
\end{figure}

Due to space restrictions in the ISAC experimental hall, the MEBT section's magnetic quadrupoles are positioned in close proximity, shown in Figure \ref{fig:mebt_lattice_spacing}. This raised concerns about possible current-dependent interference effects between adjacent lenses. Specifically, we examined whether the quadrupoles altered each others the effective magnetic lengths or introduce nonlinearities not captured by single-lens parametrizations.

To assess this, additional {\tt Opera-3D} simulations were carried out, modelling adjacent quadrupoles at representative spacing from the MEBT layout. The resulting field maps confirmed that while the fringe fields of neighboring quadrupoles do overlap, no interference effects are observed. The effective field gradients, integrated lengths, and fringe field shapes remained unchanged compared to those obtained for isolated lenses.


This outcome reflects the linear nature of Maxwell's equations: Magnetic fields from individual lenses superimpose without modifying each other's structure. As a result, overlapping regions can be treated as additive, and the optics of the system can still be described in terms of discrete, non-interacting elements.

\section{Tune Simulations}

The original design tune simulations were performed using {\tt Trace-3D}\cite{uriot2014tracewin}, without fringe field effects. These simulations assumed quadrupoles with a 2.6\,cm aperture radius and an effective hard-edge length of 18\,cm. This section investigates the effects of fringe fields upon the MEBT corner's tune.

\subsection{Addition of Fringe Fields}

\begin{figure}[!b]
  \centering
  \includegraphics[width=\linewidth]{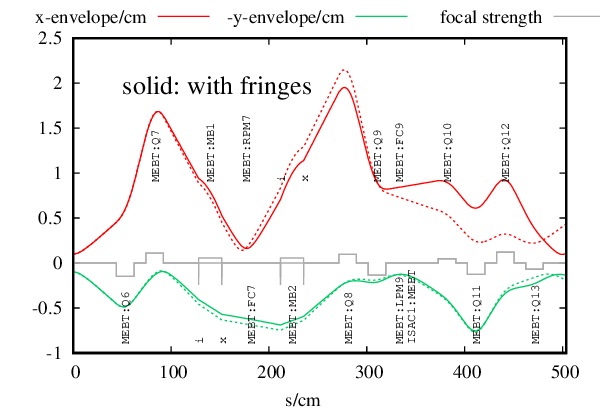}\\
  \includegraphics[width=\linewidth]{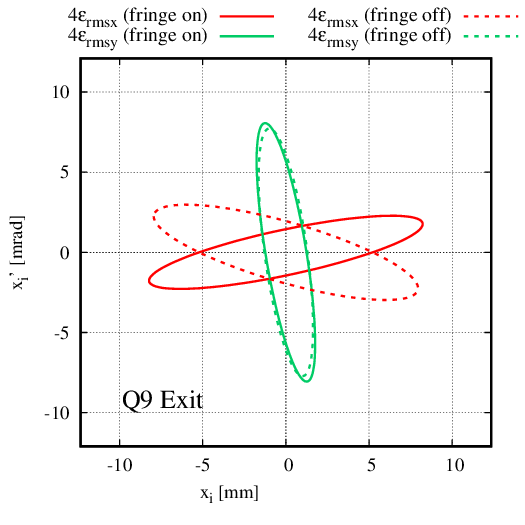}
\caption{
\textbf{Top:} Comparison of calculated beam envelopes in the MEBT corner with and without the inclusion of quadrupole fringe fields. Neglecting fringe effects introduces mismatch in both transverse planes, with a pronounced impact in the horizontal plane. 
\textbf{Bottom:} Transverse phase space distributions at the exit of Q9, where solid ellipses include fringe fields and dashed ellipses do not. The resulting shift in phase advance and envelope orientation highlights the necessity of incorporating fringe field contributions in high-precision transport beamline modeling.
}
  \label{fig:mebtTuneCompare}
\end{figure}

To evaluate the impact of the fringe field optics, two simulations were first performed. One employed the {\tt Opera-3D} derived hard-edge quadrupole model with associated fringe fields from Table~\ref{tab:quadfits}, while the other reproduced the original simulation conditions, omitting fringe field contributions. Figure~\ref{fig:mebtTuneCompare} shows the resulting MEBT corner beam envelopes for both cases, starting from the waist at the stripping foil, where the transverse tune frame is rotated by 45$^\circ$. This produces a considerable discrepancy at the start of the DTL.

To date, quadrupole setpoints used for machine operation have been obtained from fringe-field-free simulations and applied as initial conditions during linac tuning. In practice, this approach initially yields low beam transmission, requiring operators to manually adjust the quadrupole gradients in the MEBT corner to achieve high throughput. Figure~\ref{fig:mebtTuneCompare} shows the operational tune which is obtained from the manual quadrupole adjustment procedure, simulated in {\tt TRANSOPTR}, both with and without fringe field optics represented. Note $x$-broadening of the beam in the rebuncher RF cavity. When fringe fields are omitted, the $x$-envelopes into DTL are diverging, instead of converging.

While the simulation without fringe fields suggests a mismatch at injection, the same quadrupole setpoints, when simulated in an updated {\tt TRANSOPTR} model with the {\tt Opera-3D} fringe fields applied, the tune is shown to achieve the required waist at DTL IH-Tank 1.

\subsection{Comparison of Fringe Field Cases}

\begin{figure}[!htpb]
  \centering
  \includegraphics[width=\linewidth]{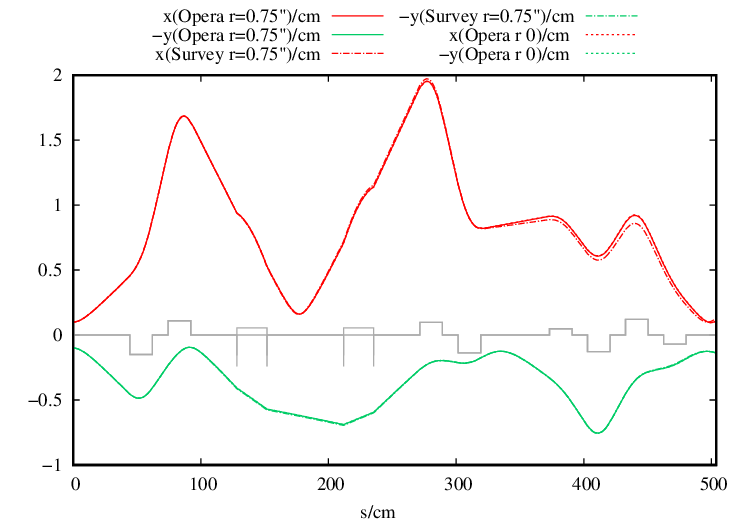}
  \caption{Comparison of transverse envelopes through the MEBT corner for the three fringe field fits listed in Table~\ref{tab:quadfits}, all using identical quadrupole current setpoints.}
  \label{fig:fringecases_comparison}
\end{figure}

A comparison is made between the three different fringe field cases summarized in Table~\ref{tab:quadfits}, illustrating the different effects of the fringe fields on the MEBT corner's tune. All three simulations are performed using identical quadrupole current setpoints. Figure~\ref{fig:fringecases_comparison} shows the evolution of the transverse beam envelopes for each case through the MEBT corner and into the DTL. 

\begin{figure*}[htpb]
  \centering
  \includegraphics[width=\textwidth]{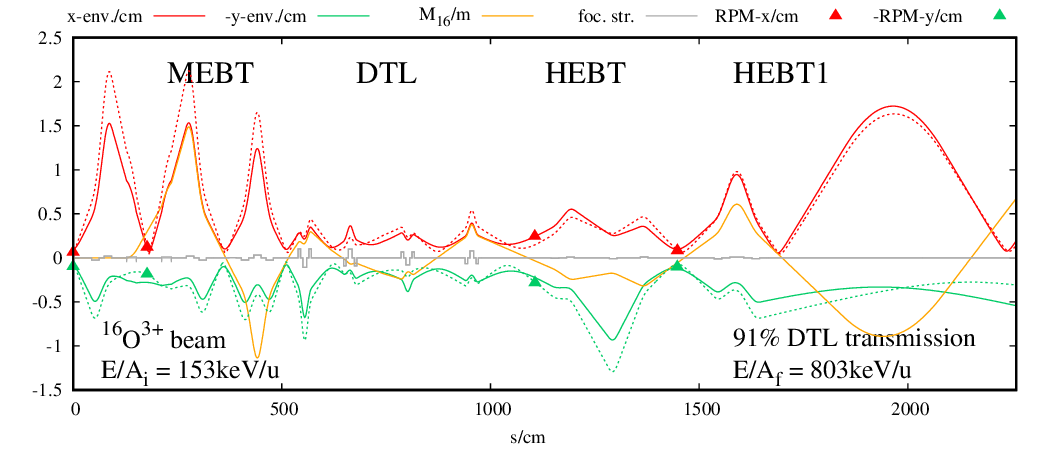}
  \caption{Comparison between the {\tt TRANSOPTR} autofocus tune (solid line) and the on-line measurements (points) for a
           $^{16}$O$^{3+}$ beam injected into the DTL at 153\,keV/u, producing the reconstructed envelopes through DTL (dotted lines).  
           The autofocus solution was calculated in seconds and loaded directly to the control
           system; the measured envelopes confirm the model within experimental uncertainty.}
  \label{fig:newOptics}
\end{figure*}
The variation in envelope sizes is small and produces a negligible change in injected beam size or divergence angle. This suggests that parallel modelling is not expected to be sensitive to the particular fringe field case chosen from Table \ref{tab:quadfits} to produce a matched beam DTL. 

For the remainder of this work, the {\tt Opera-3D} r$\approx$0 case is used.

\subsection{On-Line Tuning}

An $^{16}$O$^{3+}$ beam with an energy of 803\,keV/u was transported through the ISAC-I post-accelerator. The beam tune which was used on-line, including the MEBT corner, DTL quadrupole lattice, and the high-energy beam transport (HEBT) section, was computed from the {\tt TRANSOPTR} model\cite{TRI-BN-19-02,TRI-BN-20-08,TRI-BN-19-06}. The resulting optics solution was then applied in the accelerator control system, and the beam was threaded through the machine using corrective steerers without the need for manual adjustment of the quadrupoles for transmission optimization\cite{hassan2025strategybayesianoptimizedbeam}.

Following RF tuning to accelerate beam to 803\,keV/u, measurements of the transverse beam profiles were conducted to evaluate the agreement between the model and beam, shown in Figure \ref{fig:newOptics}. The on-line transverse phase–space distribution was determined by measuring the beam‐profile sizes\cite{TRI-BN-21-11} obtained from the rotary profile monitors (RPMs); four RPMs are available from MEBT corner to HEBT energy diagnostic station. During the measurement the IH accelerating structures were phased close to their optimum energy-gain condition, rendering them transversely equivalent to a drift, i.e.\ RF focal effects on ($x,x'$) and ($y,y'$) are minimised \cite{shelbaya2021autofocus}.

The initial distribution at the MEBT foil is fit using six free Twiss parameters, $(\alpha_x,\beta_x,\varepsilon_x,\alpha_y,\beta_y,\varepsilon_y)$, and is constrained by the eight measured 2~rms profile sizes from the RPMs. The independently measured 2~rms energy spread is used as an input to set the magnitude of transverse–longitudinal (T–L) coupling terms in the simulation.


\section{Summary and Outlook}

Previously, the TRIUMF-developed autofocus algorithm—used for automated ISAC tuning—excluded the MEBT section due to persistent discrepancies between model predictions and optimal operating conditions. It is now understood that these discrepancies stem from unmodeled fringe field effects. With the updated quadrupole model, it is possible to compute complete linac tunes, including the MEBT corner, using {\tt TRANSOPTR}, without requiring post hoc adjustment of quadrupole gradients. As a result, operators can focus tuning efforts on RF phase and beam alignment.

A systematic study of the ISAC-I medium-energy beam transport section has revealed a pronounced tune discrepancy that impacts operational reliability, when fringe field effects are not included. This sensitivity arises from a combination of factors: The use of large, low-gradient quadrupoles together with the lattice produces a tune with high sensitivity to small field errors

To address these limitations, detailed magnetic modelling of the quadrupoles was carried out using Opera-3D. The extracted fringe-field integrals were incorporated into the {\tt TRANSOPTR} beam optics model, resulting in an upgraded representation of the accelerator's lattice. This enhanced model shows excellent agreement with measured data and can now predict high-transmission beam tunes through the DTL without the need for quadrupole retuning.

As a result, reliable beam transport can be achieved by directly applying the modelled optics to the control system, eliminating the need for iterative tuning. This advance enables parallel, simulation-based tuning of the accelerator, significantly simplifying beamline operation and improving efficiency for future ISAC experiments.

\section{Acknowledgements}

Thanks to the RIB operations group for support during this work. This work has been funded by NSERC grant SAPPJ-2023-00038. The TRIUMF campus is located on the traditional, ancestral, and unceded territory of the Musqueam people.

\bibliographystyle{elsarticle-num-names} 

\begin{thebibliography}{27}
\expandafter\ifx\csname natexlab\endcsname\relax\def\natexlab#1{#1}\fi
\providecommand{\url}[1]{\texttt{#1}}
\providecommand{\href}[2]{#2}
\providecommand{\path}[1]{#1}
\providecommand{\DOIprefix}{doi:}
\providecommand{\ArXivprefix}{arXiv:}
\providecommand{\URLprefix}{URL: }
\providecommand{\Pubmedprefix}{pmid:}
\providecommand{\doi}[1]{\href{http://dx.doi.org/#1}{\path{#1}}}
\providecommand{\Pubmed}[1]{\href{pmid:#1}{\path{#1}}}
\providecommand{\bibinfo}[2]{#2}
\ifx\xfnm\relax \def\xfnm[#1]{\unskip,\space#1}\fi
\bibitem[{Kunz et~al.(2014)Kunz, Andreoiu, Bricault, Dombsky, Lassen, Teigelh{\"o}fer, Heggen, and Wong}]{kunz2014nuclear}
\bibinfo{author}{P.~Kunz}, \bibinfo{author}{C.~Andreoiu}, \bibinfo{author}{P.~Bricault}, \bibinfo{author}{M.~Dombsky}, \bibinfo{author}{J.~Lassen}, \bibinfo{author}{A.~Teigelh{\"o}fer}, \bibinfo{author}{H.~Heggen}, \bibinfo{author}{F.~Wong},
\newblock \bibinfo{title}{Nuclear and in-source laser spectroscopy with the isac yield station},
\newblock \bibinfo{journal}{Review of Scientific Instruments} \bibinfo{volume}{85} (\bibinfo{year}{2014}) \bibinfo{pages}{053305}.
\bibitem[{Laxdal et~al.(1997)Laxdal, Bricault, Reis, and Gorelov}]{Laxdal:1997ge}
\bibinfo{author}{R.~E. Laxdal}, \bibinfo{author}{P.~G. Bricault}, \bibinfo{author}{T.~Reis}, \bibinfo{author}{D.~V. Gorelov},
\newblock \bibinfo{title}{{A separated function drift-tube linac for the ISAC project at TRIUMF}},
\newblock \bibinfo{journal}{Conf. Proc. C} \bibinfo{volume}{970512} (\bibinfo{year}{1997}) \bibinfo{pages}{1194--1196}.
\bibitem[{Shelbaya(2019)}]{TRI-BN-19-02}
\bibinfo{author}{O.~Shelbaya}, \bibinfo{title}{{TRANSOPTR Implementation of the MEBT Beamline}}, \bibinfo{type}{Technical Report} \bibinfo{number}{TRI-BN-19-02}, TRIUMF, \bibinfo{year}{2019}.
\bibitem[{Laxdal et~al.(2002)Laxdal, Pasini, Root et~al.}]{laxdal2002beam}
\bibinfo{author}{R.~Laxdal}, \bibinfo{author}{M.~Pasini}, \bibinfo{author}{L.~Root}, et~al.,
\newblock \bibinfo{title}{Beam dynamics design study and beam commissioning of the isac two frequency chopper},
\newblock \bibinfo{journal}{LINAC2002, Gyeongju, Korea}  (\bibinfo{year}{2002}) \bibinfo{pages}{407}.
\bibitem[{Shelbaya(2022)}]{TRI-BN-22-29}
\bibinfo{author}{O.~Shelbaya}, \bibinfo{title}{{Beam Dynamics Study of ISAC-MEBT}}, \bibinfo{type}{Technical Report} \bibinfo{number}{TRI-BN-22-29}, {TRIUMF}, \bibinfo{year}{2022}.
\bibitem[{Shelbaya et~al.(2021)Shelbaya, Angus, Baartman, Jung, Kester, Kiy, Planche, and R\"adel}]{shelbaya2021autofocus}
\bibinfo{author}{O.~Shelbaya}, \bibinfo{author}{T.~Angus}, \bibinfo{author}{R.~Baartman}, \bibinfo{author}{P.~M. Jung}, \bibinfo{author}{O.~Kester}, \bibinfo{author}{S.~Kiy}, \bibinfo{author}{T.~Planche}, \bibinfo{author}{S.~D. R\"adel},
\newblock \bibinfo{title}{Autofocusing drift tube linac envelopes},
\newblock \bibinfo{journal}{Phys. Rev. Accel. Beams} \bibinfo{volume}{24} (\bibinfo{year}{2021}) \bibinfo{pages}{124602}. \DOIprefix\doi{10.1103/PhysRevAccelBeams.24.124602}.
\bibitem[{Merminga et~al.(2011)Merminga, Ames, Baartman, Bricault, Bylinski, Chao, Dawson, Kaltchev, Koscielniak, Laxdal et~al.}]{merminga2011ariel}
\bibinfo{author}{L.~Merminga}, \bibinfo{author}{F.~Ames}, \bibinfo{author}{R.~Baartman}, \bibinfo{author}{P.~Bricault}, \bibinfo{author}{Y.~Bylinski}, \bibinfo{author}{Y.~Chao}, \bibinfo{author}{R.~Dawson}, \bibinfo{author}{D.~Kaltchev}, \bibinfo{author}{S.~Koscielniak}, \bibinfo{author}{R.~Laxdal}, et~al.,
\newblock \bibinfo{title}{Ariel: Triumf’s advanced rare isotope laboratory},
\newblock \bibinfo{journal}{WEOBA001, IPAC} \bibinfo{volume}{11} (\bibinfo{year}{2011}) \bibinfo{pages}{1917--1918}.
\bibitem[{Bagger et~al.(2018)}]{Bagger:IPAC2018-MOXGB2}
\bibinfo{author}{J.~Bagger}, et~al.,
\newblock \bibinfo{title}{{ARIEL} at {TRIUMF}: {S}cience and {T}echnology},
\newblock in: \bibinfo{booktitle}{Proc. 9th International Particle Accelerator Conference (IPAC'18), Vancouver, BC, Canada, April 29-May 4, 2018}, number~\bibinfo{number}{9} in \bibinfo{series}{International Particle Accelerator Conference}, \bibinfo{publisher}{JACoW Publishing}, \bibinfo{address}{Geneva, Switzerland}, \bibinfo{year}{2018}, pp. \bibinfo{pages}{6--11}.
\bibitem[{Dilling et~al.(2013)Dilling, Kr{\"u}cken, and Merminga}]{dilling2013ariel}
\bibinfo{author}{J.~Dilling}, \bibinfo{author}{R.~Kr{\"u}cken}, \bibinfo{author}{L.~Merminga},
\newblock \bibinfo{title}{Ariel overview},
\newblock in: \bibinfo{booktitle}{ISAC and ARIEL: The TRIUMF Radioactive Beam Facilities and the Scientific Program}, \bibinfo{publisher}{Springer}, \bibinfo{year}{2013}, pp. \bibinfo{pages}{253--262}.
\bibitem[{Wang et~al.(2021)Wang, Bagri, Macdonald, Kiy, Jung, Shelbaya, Planche, Fedorko, Baartman, and Kester}]{wang2021accelerator}
\bibinfo{author}{D.~Y. Wang}, \bibinfo{author}{H.~Bagri}, \bibinfo{author}{C.~Macdonald}, \bibinfo{author}{S.~Kiy}, \bibinfo{author}{P.~Jung}, \bibinfo{author}{O.~Shelbaya}, \bibinfo{author}{T.~Planche}, \bibinfo{author}{W.~Fedorko}, \bibinfo{author}{R.~Baartman}, \bibinfo{author}{O.~Kester},
\newblock \bibinfo{title}{Accelerator tuning with deep reinforcement learning},
\newblock in: \bibinfo{booktitle}{Workshop on Machine Learning and the Physical Sciences}, \bibinfo{year}{2021}, pp. \bibinfo{pages}{1,11}.
\bibitem[{Ghelfi et~al.(2025)Ghelfi, Katrusiak, Baartman, Fedorko, Kester, Kogler~Anele, Shelbaya, and Tanyer}]{ghelfi2025bayesian}
\bibinfo{author}{E.~Ghelfi}, \bibinfo{author}{A.~Katrusiak}, \bibinfo{author}{R.~Baartman}, \bibinfo{author}{W.~Fedorko}, \bibinfo{author}{O.~Kester}, \bibinfo{author}{G.~Kogler~Anele}, \bibinfo{author}{O.~Shelbaya}, \bibinfo{author}{D.~Tanyer},
\newblock \bibinfo{title}{Bayesian optimization for ion beam centroid correction},
\newblock \bibinfo{journal}{Review of Scientific Instruments} \bibinfo{volume}{96} (\bibinfo{year}{2025}).
\bibitem[{Shelbaya et~al.(2024)Shelbaya, Baartman, Braun, Jung, Kester, Planche, Podlech, and Rädel}]{shelbaya2024tuning}
\bibinfo{author}{O.~Shelbaya}, \bibinfo{author}{R.~Baartman}, \bibinfo{author}{P.~Braun}, \bibinfo{author}{P.~M. Jung}, \bibinfo{author}{O.~Kester}, \bibinfo{author}{T.~Planche}, \bibinfo{author}{H.~Podlech}, \bibinfo{author}{S.~D. Rädel},
\newblock \bibinfo{title}{Tuning methods for multigap drift tube linacs},
\newblock \bibinfo{journal}{Review of Scientific Instruments} \bibinfo{volume}{95} (\bibinfo{year}{2024}) \bibinfo{pages}{033302}. \DOIprefix\doi{10.1063/5.0191603}.
\bibitem[{Marchetto and Laxdal(2014)}]{marchetto2014high}
\bibinfo{author}{M.~Marchetto}, \bibinfo{author}{R.~Laxdal},
\newblock \bibinfo{title}{High energy beam lines},
\newblock \bibinfo{journal}{{ISAC and ARIEL: The TRIUMF Radioactive Beam Facilities and the Scientific Program}}  (\bibinfo{year}{2014}) \bibinfo{pages}{99--109}.
\bibitem[{Shelbaya et~al.(2019)Shelbaya, Baartman, and Kester}]{shelbaya2019fast}
\bibinfo{author}{O.~Shelbaya}, \bibinfo{author}{R.~Baartman}, \bibinfo{author}{O.~Kester},
\newblock \bibinfo{title}{Fast radio frequency quadrupole envelope computation for model based beam tuning},
\newblock \bibinfo{journal}{Physical Review Accelerators and Beams} \bibinfo{volume}{22} (\bibinfo{year}{2019}) \bibinfo{pages}{114602}.
\bibitem[{Laxdal and Marchetto(2014)}]{laxdal2014isac}
\bibinfo{author}{R.~Laxdal}, \bibinfo{author}{M.~Marchetto},
\newblock \bibinfo{title}{{The ISAC post-accelerator}},
\newblock \bibinfo{journal}{Hyperfine Interactions} \bibinfo{volume}{225} (\bibinfo{year}{2014}) \bibinfo{pages}{79--97}.
\bibitem[{Marchetto et~al.(2010)Marchetto, Aoki, Langton, Laxdal, Rawnsley, Richards et~al.}]{marchetto2010isac}
\bibinfo{author}{M.~Marchetto}, \bibinfo{author}{J.~Aoki}, \bibinfo{author}{K.~Langton}, \bibinfo{author}{R.~Laxdal}, \bibinfo{author}{W.~Rawnsley}, \bibinfo{author}{J.~Richards}, et~al.,
\newblock \bibinfo{title}{The isac-ii current monitor system},
\newblock \bibinfo{journal}{Proc. LINAC10, Tsukuba Japan}  (\bibinfo{year}{2010}).
\bibitem[{Laxdal et~al.(1997)Laxdal, Bricault, Reis, and Gorelov}]{laxdal1997separated}
\bibinfo{author}{R.~Laxdal}, \bibinfo{author}{P.~Bricault}, \bibinfo{author}{T.~Reis}, \bibinfo{author}{D.~Gorelov},
\newblock \bibinfo{title}{A separated function drift-tube linac for the isac project at triumf},
\newblock in: \bibinfo{booktitle}{Proceedings of the 1997 Particle Accelerator Conference (Cat. No. 97CH36167)}, volume~\bibinfo{volume}{1}, \bibinfo{organization}{IEEE}, \bibinfo{year}{1997}, pp. \bibinfo{pages}{1194--1196}.
\bibitem[{Shelbaya and Baartman(2019)}]{TRI-BN-19-18}
\bibinfo{author}{O.~Shelbaya}, \bibinfo{author}{R.~Baartman}, \bibinfo{title}{{Langevin-Like DTL Triplet BI Fits and Analysis of Transverse DTL Tuning Difficulties}}, \bibinfo{type}{Technical Report} \bibinfo{number}{TRI-BN-19-18}, TRIUMF, \bibinfo{year}{2019}.
\bibitem[{Heighway and Hutcheon(1981)}]{heighway1981transoptr}
\bibinfo{author}{E.~Heighway}, \bibinfo{author}{R.~Hutcheon},
\newblock \bibinfo{title}{{{TRANSOPTR}—A second order beam transport design code with optimization and constraints}},
\newblock \bibinfo{journal}{Nuclear Instruments and Methods in Physics Research} \bibinfo{volume}{187} (\bibinfo{year}{1981}) \bibinfo{pages}{89--95}. \URLprefix \url{https://doi.org/10.1016/0029-554X(81)90474-2}. \DOIprefix\doi{10.1016/0029-554X(81)90474-2}.
\bibitem[{Baartman(2022)}]{TRI-BN-22-08}
\bibinfo{author}{R.~Baartman}, \bibinfo{title}{{ TRANSOPTR Reference Manual }}, \bibinfo{type}{Technical Report} \bibinfo{number}{TRI-BN-22-08}, TRIUMF, \bibinfo{year}{2022}.
\bibitem[{Matsuda and Wollnik(1972)}]{matsuda1972third}
\bibinfo{author}{H.~Matsuda}, \bibinfo{author}{H.~Wollnik},
\newblock \bibinfo{title}{Third order transfer matrices for the fringing field of magnetic and electrostatic quadrupole lenses},
\newblock \bibinfo{journal}{Nuclear Instruments and Methods} \bibinfo{volume}{103} (\bibinfo{year}{1972}) \bibinfo{pages}{117--124}.
\bibitem[{Baartman and Kaltchev(2007)}]{baartman2007short}
\bibinfo{author}{R.~Baartman}, \bibinfo{author}{D.~Kaltchev},
\newblock \bibinfo{title}{Short quadrupole parametrization},
\newblock in: \bibinfo{booktitle}{Particle Accelerator Conference, 2007 PAC.\ IEEE}, \bibinfo{organization}{IEEE}, \bibinfo{year}{2007}, pp. \bibinfo{pages}{3229--3231}.
\bibitem[{Uriot and Pichoff(2014)}]{uriot2014tracewin}
\bibinfo{author}{D.~Uriot}, \bibinfo{author}{N.~Pichoff},
\newblock \bibinfo{title}{Tracewin},
\newblock \bibinfo{journal}{CEA Saclay} \bibinfo{volume}{596} (\bibinfo{year}{2014}).
\bibitem[{Shelbaya(2020)}]{TRI-BN-20-08}
\bibinfo{author}{O.~Shelbaya}, \bibinfo{title}{{The TRANSOPTR Model of the ISAC Drift Tube Linear Accelerator - Part I: Longitudinal Verification}}, \bibinfo{type}{Technical Report} \bibinfo{number}{TRI-BN-20-08}, TRIUMF, \bibinfo{year}{2020}.
\bibitem[{Shelbaya(2019)}]{TRI-BN-19-06}
\bibinfo{author}{O.~Shelbaya}, \bibinfo{title}{{TRANSOPTR Implementation of the HEBT Beamlines}}, \bibinfo{type}{Technical Report} \bibinfo{number}{TRI-BN-19-06}, TRIUMF, \bibinfo{year}{2019}.
\bibitem[{Hassan et~al.(2025)Hassan, Shelbaya, Fedorko, Planche, and Kester}]{hassan2025strategybayesianoptimizedbeam}
\bibinfo{author}{O.~Hassan}, \bibinfo{author}{O.~Shelbaya}, \bibinfo{author}{W.~Fedorko}, \bibinfo{author}{T.~Planche}, \bibinfo{author}{O.~Kester},
\newblock \bibinfo{title}{Strategy for bayesian optimised beam steering at triumf-isac's mebt and hebt beamlines},
\newblock \bibinfo{journal}{Journal of Instrumentation} \bibinfo{volume}{20} (\bibinfo{year}{2025}) \bibinfo{pages}{T07005}. \URLprefix \url{https://dx.doi.org/10.1088/1748-0221/20/07/T07005}. \DOIprefix\doi{10.1088/1748-0221/20/07/T07005}.
\bibitem[{Wu(2021)}]{TRI-BN-21-11}
\bibinfo{author}{K.~Wu}, \bibinfo{title}{{Profile Monitor Classification using Random Forest Classifier}}, \bibinfo{type}{Technical Report} \bibinfo{number}{TRI-BN-21-11}, TRIUMF, \bibinfo{year}{2021}.

\end{thebibliography}
\providecommand{\noopsort}[1]{}\providecommand{\singleletter}[1]{#1}%

\end{document}